\documentclass[%
 reprint,
showpacs,preprintnumbers,
 amsmath,amssymb,
 aps,
pre,
floatfix,
]{revtex4-1}
\usepackage{graphicx}
\usepackage{dcolumn}
\usepackage{bm}

\begin{document}


\title{Flow of granular matter in a silo with multiple exit orifices:
  Jamming to mixing}

\author{Sandesh Kamath} \affiliation{Chemical Engineering Division,
  National Chemical Laboratory, Pune 411008 India} 
\author{Amit Kunte} \affiliation{Chemical Engineering Division,
  National Chemical Laboratory, Pune 411008 India} 
\author{Pankaj Doshi} \email{p.doshi@ncl.res.in} \affiliation{Chemical
  Engineering Division, National Chemical Laboratory, Pune 411008
  India}
\author{Ashish V. Orpe} \email{av.orpe@ncl.res.in}
\affiliation{Chemical Engineering Division, National Chemical
  Laboratory, Pune 411008 India}

\date{\today}

\begin{abstract}
  We investigate the mixing characteristics of dry granular material
  while draining down a silo with multiple exit orifices. The mixing
  in the silo, which otherwise consists of noninteracting stagnant
  and flow regions, is observed to improve significantly when the flow
  through specific orifices is stopped intermittently. This
  momentary stoppage of flow through the orifice is either controlled
  manually or is chosen by the system itself when the orifice width is
  small enough to cause spontaneous jamming and unjamming. We observe
  that the overall mixing behavior shows a systematic dependence on
  the frequency of closing and opening of specific orifices. In
  particular, the silo configuration employing random jamming and
  unjamming of any of the orifices shows early evidence of chaotic
  mixing. When operated in a multipass mode, the system exhibits a
  practical and efficient way of mixing particles.
\end{abstract}

\maketitle

\section{\label{intro}Introduction}

Flow of granular media through a hopper or silo has a ubiquitous
presence in several industrial applications and has been investigated
over a long time. The flow consists of particle drainage under the
influence of gravity and comprises an accelerating section close to
the exit orifice and a slow plug region high above the
orifice. Various modeling approaches have been used to describe the
entire flow behavior, viz., the kinematic model~\cite{nedder79}, the
void model~\cite{mullins79}, the spot model~\cite{choi05}, the
frictional-cosserat model~\cite{mohan99} and many others. The output
flow rate from the silo is known to follow the famous Beverloo
correlation over a wide range of system parameters~\cite{anand08}. The
system is, however, more known, or quite notorious, for its uncanny
ability to jam suddenly due to the formation of highly stable arches
at the exit orifice~\cite{tang11,janda08,tewari13} which can cause
many problems in industrial operations.

Typically, the particles in a silo are densely packed, and when they
flow, the motion is quite slow in the transverse direction as compared
with the downward direction. Additionally, the system is comprised of
stagnant zones, particularly in a rectangular bottom silo, which does
not show significant particle motion throughout the silo drainage. All
these characteristics add up to result in little or almost non
existent mixing behavior, as evidenced through the advection dominated
particle dynamics in the downward flow direction compared with
transverse diffusion~\cite{choi04}.

In this work, we try to ascertain whether a silo can be used for
efficient granular matter mixing. We seek motivation from our recent
work~\cite{kunte14} wherein we observed the effect of nonlocal flow
behavior on the jamming characteristics in a hopper with multiple exit
orifices. We showed that the fluctuations emanating due to flow from
one orifice of a silo can cause another jammed orifice located as far
as $30$ particle diameters to spontaneously unjam and flow. When the
two orifices are very close to each other (less than $3$ particle
diameters), the jamming of orifices can be prevented due to mutually
interacting arches over both orifices~\cite{mondal14}. We extend this
concept of jamming and unjamming to demonstrate that, in a silo with
multiple exit orifices, the particles above a jammed orifice can pass
through the other adjacent flowing orifice, resulting in substantial
cross flow.  The result is the enhanced mixing behavior observed in
the system. We show this behavior using two different ways of creating
the desired cross flow in the system, viz., (i) closing (or jamming)
and opening (or unjamming) of an orifice in a systematic, controlled
manner and (ii) allowing the orifices in the silo to jam and unjam
randomly depending on their interorifice distance and orifice width.
It is to be noted that this enhanced mixing behavior using both
schemes is enabled through flow re-arrangements created noninvasively
and hitherto unobserved in a silo system. Such enhanced mixing
behavior will be of immense importance to several applications wherein
the silo acts as a feeder to a process requiring uniformly mixed
material or is a part of a larger integrated system and requires
efficient mixing throughout the height as in a nuclear pebble bed
reactor.  We explore this enhanced mixing behavior in detail by using
discrete element method (DEM) simulations of soft particles. The
particle positions obtained throughout the course of simulation are
used to evaluate the spatial and temporal dependence of the mixing
properties, measured as degree of mixing.

\section{\label{meth}Methodology}

The DEM simulations are carried out using the Large Atomic-Molecular
Massively Parallel Simulator {\sc(lammps)} developed at Sandia
National laboratories~\cite{plimp95,lammps}. The simulation employs
Hookean force between two contacting particles described in detail
elsewhere~\cite{rycroft09}. All the simulation parameters are the same
as used in the systematic study of silo flows carried out
previously~\cite{rycroft09}, except for a higher normal elastic
constant ($k_n = 2 \times 10^6 mg/d$) which corresponds to a more
stiff particle. The interparticle friction coefficient ($\mu$) is
varied from $0.2$ to $0.8$ with no qualitative difference between the
results. Here, we report the results obtained for $\mu = 0.5$.

\begin{figure}
  \centering\includegraphics[width=0.75\linewidth]{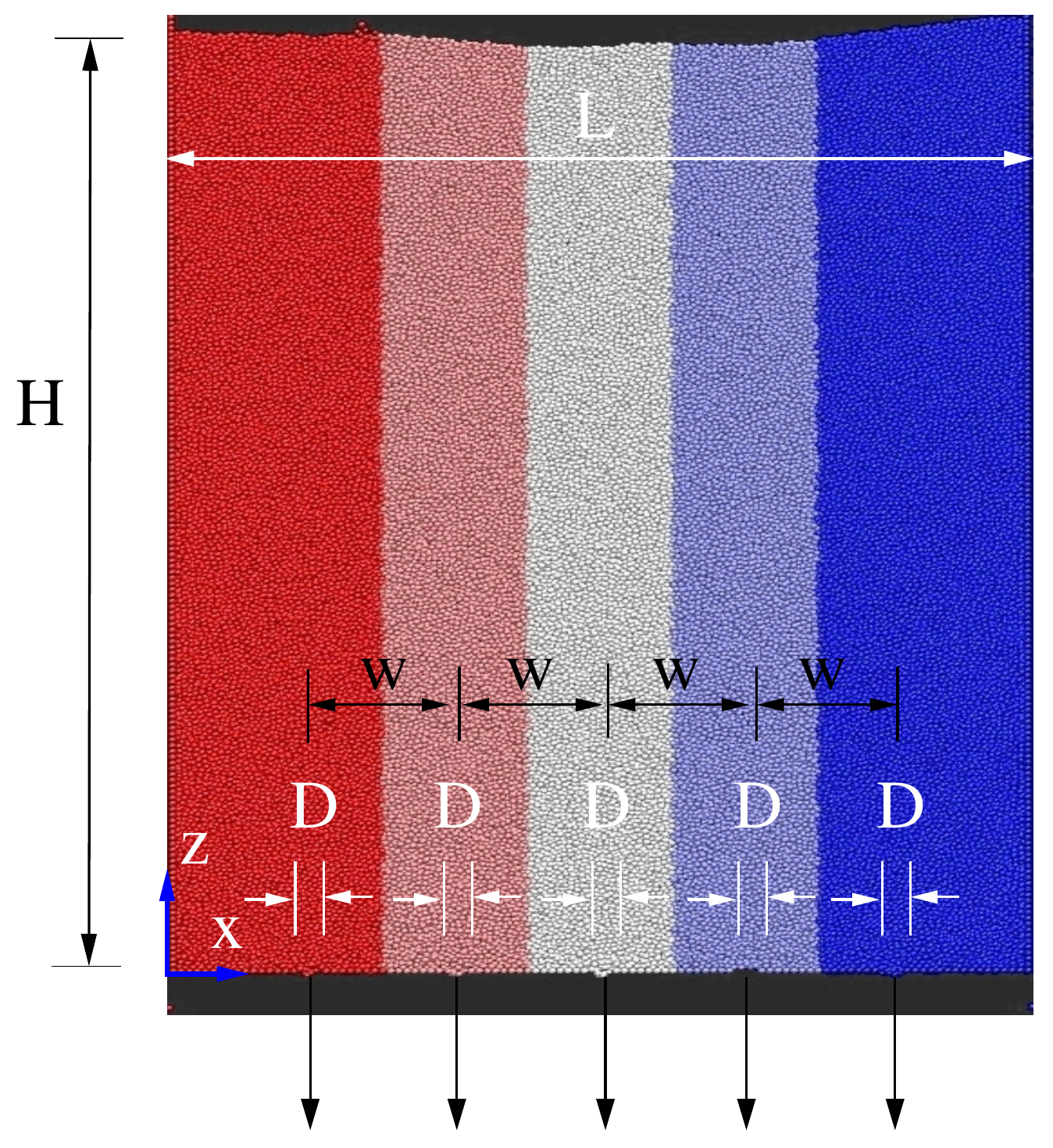}
  \caption{(Color online) Sample snapshot of the initial silo before
    the flow ensues. All the particles in the system are of same size
    and polydispersity. The color code is employed to visualize and
    identify particles over different orifices and in different
    regions.}
  \label{fig1}
\end{figure}

The simulation geometry (see Fig.~\ref{fig1}) consists of a
rectangular, flat-bottomed, silo of height $H$, width $L$, and
thickness $3d$, where $d \pm 0.15d$ is the particle diameter with a
uniform size distribution. The side and bottom walls are created by
freezing the largest sized particles so that their translational and
angular velocities are kept zero throughout the simulation run. The
silo has five orifices, each of width $D$ and separated by the
interorifice distance $w$. Periodic boundaries are used in the $y$
direction (in and out of the paper), which represents an infinitely
long silo in $y$ direction. The silo is filled by using the
sedimentation method as suggested previously~\cite{landry03} in which
a dilute packing of nonoverlapping particles is created in a
simulation box and allowed to settle under the influence of
gravity. The simulation is run for a significant time so that the
kinetic energy per particle is less than $10^{-8} mgd$ resulting in a
quiescent packing of height $H$ in the silo. All the particles in the
silo are the same type and size, but colored differently to create
five different vertical bands, each centered above one of the five
orifices.

The silo system is operated in two ways, viz., (i) single-pass and (ii)
multipass. In a single-pass system, the particles traverse down the
entire height of the silo only once before exiting through the orifice
located at the bottom. The height ($H$) of each column of (colored)
particles is maintained constant at $160d$ by inserting new particles
of same color at the free surface equal in number to those which have
exited from the system per unit time. The free surface remains nearly
flat during the entire simulation run. This mode of operation
represents an infinitely tall silo or a silo with a finite height
continuously receiving a fresh feed of material. In contrast, in a
multi-pass system every particle traverses down the entire silo height
several times during the simulation run. The same particles are
re-inserted at the free surface and at the same horizontal position
from where they exited the system. The average fill height ($H$)
during the simulation run remains constant at $80d$ while the free
surface can exhibit significant slope, as discussed later. This mode
of operation represents a batch system with the particles subjected to
a particular flow mechanism repeatedly. A vertical cascade of silos,
wherein the material exiting from one silo would enter at the free
surface of another silo located exactly below it, would represent such
a batch system. 

The flow through the silo system is controlled using two protocols.
In one scenario, the second and fourth orifice from either sidewall
are opened and closed alternately while keeping the remaining three
(first, third, and fifth) orifices open throughout. The orifice
(either second or fourth) is kept open, while keeping the fourth or
second closed, respectively, for a time ($\Delta t$) within which the
mean distance drop of the particles in the silo is approximately $T =
\langle v_{z} \rangle \Delta t /d$, where $\langle v_{z} \rangle$ is
the absolute magnitude of the downward velocity of particles averaged
across the silo width. The mixing characteristics are determined for
various values of $T$. The simulations for each value of $T$ are
continued until the total number $n$ of instances of closing and
opening of orifices reach fourteen (both second and fourth orifices
opened and closed seven times each). The total run-time $t$ of the
simulations is, thus, different for each value of $T$ . Here time $t$
is measured in terms of unit $\tau = \sqrt{d/g}$, which is the natural
timescale of simulations and $g$ is the acceleration due to
gravity. To stop the flow through an orifice, the particles in a
region $3d \times 5d$ just above the orifice are suddenly
frozen. Similarly, to re-initiate the flow, these frozen particles are
allowed to fall freely under gravity. In an experimental system, this
is equivalent to a rapid closing and opening of the orifice valve,
respectively, to stop and restart the flow.  The width ($D$) of each
orifice is $3d$ which ensures a constant flow rate from the orifice as
long as the orifice is kept open. The interorifice distance ($w$),
i.e., the center-to-center distance between adjacent orifices, is kept
constant throughout at $15d$. The overall silo width ($L$), excluding
the side-wall particles, is $105d$ wherein each extreme orifice is
placed $22.5d$ away from the nearest sidewall. In the subsequent
sections, we refer to this protocol as the ``controlled mechanism''.

In another scenario, the width of all the orifices is reduced to $2d$
while the interorifice distance is maintained almost the same ($w
\approx 15d$). This change in the system causes a dramatic change in
the flow behavior. The smaller orifice width can cause jamming of
flow, only to be unjammed due to fluctuations transferred from one of
the other, albeit intermittently flowing, adjacent
orifices~\cite{kunte14}. The probability of this unjamming depends on
the distance between two adjacent orifices and the orifice width. For
a given orifice width, the closer the two orifices, the more probable
is unjamming, while this probability is very small when the orifices
are far apart. In the latter case, the fluctuations cannot traverse
that long to unjam a jammed orifice. The combination of $w$ and $D$
used here allows for this random jamming and unjamming to persist
without complete stoppage of flow and ensuring flow through at least
two orifices at any point of time. The resulting effect is that the
material above the jammed orifice can cross over to the adjacent
region, leading to mixing. However, the value of $T$ is now variable
throughout the course of simulation, which is chosen randomly by the
system on its own. Unlike the controlled-mechanism case where only a
specific orifice can close or open, in this case one or more orifices
can randomly jam or unjam. It should be noted that different
combinations of $D$ and $w$ are possible, albeit within a very limited
range, to initiate this random jamming- and unjamming-induced
flow. We, however, focus here on the fixed $D$ and $w$ mentioned above
to simply provide a flavor of the mixing dynamics arising out of this
flow scheme.  The width ($L$) of the silo, excluding the sidewall
particles, in this case is $74.5d$ wherein the extreme orifices are
placed $7.5d$ away from the nearest sidewall. We refer to this
protocol as the ``random mechanism'' in subsequent sections.

To characterize mixing in the system, a column of particles above
individual orifices are colored differently (see Fig.~\ref{fig1}). A
horizontal box of height ($h = 4 d$) and width ($l = 18 d$) and depth
$3d$ is centered exactly at a distance $4d$ above each orifice [see
Fig.~\ref{fig2}(a) and 2(b)]. The cumulative particle fraction
$\langle \phi \rangle$ for the entire silo at any time ($t$) is defined as
\begin{equation}
  \langle \phi \rangle = \frac{\sum{\phi_{i}}}{n_{o}} = \frac{\sum{N_{i}(t)/N_{i}(0)}}{n_{o}},
\end{equation}
where $i$ denotes the region above each orifice, $\phi_{i}$ is the
fraction corresponding to box ($h \times l \times 3d$) above each
orifice, and $n_{o}$ is the number of orifices. The summation is
carried over the values from all five boxes located, respectively,
over five orifices. Initially, each box ($h \times l \times 3d$)
contains particles of only one color and the number is
$N_{i}(0)$. The number of particles of that color in the same box at
different times ($t$) is $N_{i}(t)$. The cumulative
particle fraction, thus, measures the fraction of particular colored
particles in a box preserved over time ($t$). The degree of
mixing in the system is defined as
\begin{equation}
  M = \frac{1-\langle \phi \rangle}{1-1/n_{o}}.
\end{equation}
For five orifices ($n_{o} = 5$) in the system, the value of
$\langle \phi \rangle$ can vary from $1$ for ``no mixing'' to $0.2$
for ``complete mixing''. Correspondingly, the value of $M$ varies from
$0$ (no mixing) to $1$ (complete mixing).
Measurements of $M$ over
time provides steady state evolution 
of mixing behavior. Similarly, measurements in the same sized box
but located at different heights provides the spatial variation of
mixing.

\section{\label{results}{Results}}

\subsection{\label{spass}{Single pass system}}

\begin{figure}
  \centering\includegraphics[width=1.0\linewidth]{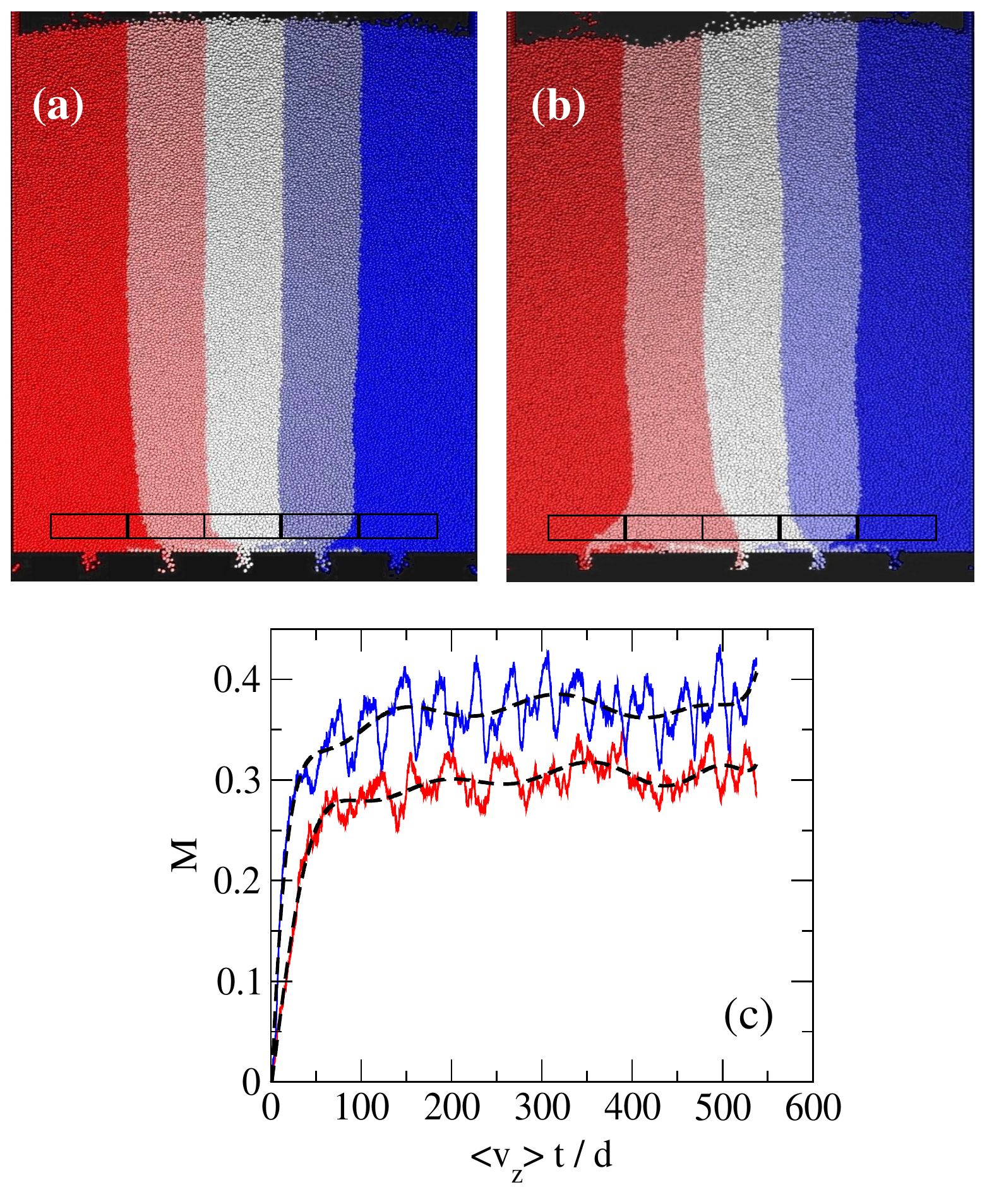}
  \caption{(Color online) Mixing patterns and degree of mixing for
    various orifice configurations. Snapshots obtained for (a) all
    orifices open ($T = 0.0$), (b) second and fourth orifices opened
    and closed alternately, keeping $T = 38.4$ at $\langle v_{z}
    \rangle t/d = 400$.  Black boxes at the bottom represents the
    region in which $M$ is calculated. (c) Variation in the degree of
    mixing ($M$) with time for two cases: red (lower curve) denotes
    the configuration in panel (a), and blue (upper curve) denotes the
    configuration in panel (b). The dashed black lines show a
    tenth-degree-polynomial fit.}
  \label{fig2}
\end{figure}

In this section, we discuss the results obtained using different
flow-controlling methods in the system through which every particle
traverses only once. We first discuss the mixing behavior for manually
controlled flow through orifices. Figure~\ref{fig2}(b) shows a
snapshot during the flow obtained for $T =38.4$, which is the largest
value considered. Also shown in Fig.~\ref{fig2}(a) is the case for $T
=0$ (i.e., all orifices open at all times). The mixing for
intermediate values of $T$ is discussed later.  More details about the
pattern evolution can be observed in the movies uploaded as
supplementary material for different cases.  The steady state
evolution of the degree of mixing is shown in Fig.~\ref{fig2}(c). For
both the cases the simulations are continued for a time duration $t$
so that the value of $M$ reaches a near constant value and the total
mean downward distance traveled ($\langle v_{z} \rangle t/d$) is the
same in both the cases.  To obtain an estimate of the steady state
value, a tenth-order polynomial is fit to the entire data set [dashed
black lines shown in Fig.~\ref{fig2}(c)]. The steady state value
($M_{s}$) is then obtained from the average of the polynomial-fit
values over a time domain within which its standard deviation is less
than $1$\%. In Fig.~\ref{fig2}(c), this time domain corresponds to
$200 <=\langle v_{z} \rangle t/d <= 600$ for both the cases. The
images shown in Figs.~\ref{fig2}(a) and 2(b) are obtained at time
instant $\langle v_{z} \rangle t/d = 400$.
 
When all orifices are kept open at all times, the system represents,
to a certain extent, five single-orifice silos of smaller widths
operating in parallel. We find particles flowing within all regions of
the silo, thus, eliminating the presence of any stagnant regions,
particularly near the bottom corners, typically observed in a
single-orifice flat-bottomed silo. It is observed that the particles
from one column (of a particular color) flow through the orifice
located exactly below the adjacent column [of another color; see
Fig.~\ref{fig2}(a)]. The reason for this is the interaction between
the spatial regions above each orifice. Note that, for a
single-orifice silo, a convergent-flow region exists just above the
orifice which extends vertically upwards up to a height approximately
equal to the width of the silo while simultaneously also extending
horizontally up to a distance approximately equal to half the silo
width. The typical downward velocity profile across the width at any
height in the convergent section has a Gaussian
shape~\cite{nedder79,choi05}. It is, thus, quite natural for the flow
fields above each orifice, in a multiple exit orifice silo, to
interact with each other in a nonlinear manner. The particles on or
near the interface of the two adjacent columns will, thus, have an
equal probability of going through either of the two adjacent
orifices, leading to cross flow and mixing of particles with $M_{s}
\approx 0.3$ [see Fig.~\ref{fig2}(c)]. The profile after reaching
steady state shows sustained oscillations about the value of $M_{s}$
obtained from the polynomial fit. The amplitude and period of these
oscillations are governed by the time-dependent inherent dynamics of
the coupling between two adjacent flow fields.

For the controlled mechanism with $T = 38.4$, more cross flow is
observed, which also extends to larger heights above the orifice [see
Fig.~\ref{fig2}(b)]. The degree of mixing evolves to a steady state
value of $0.375$ [shown as (upper) blue curve in Fig.~\ref{fig2}(c)]
which is higher than for the case $T = 0.0$.  Here as well, the value
of $M$, after reaching steady state, continues to oscillate.  The
oscillations in this case arise due to sequential closing and opening
of orifices, causing particles from a particular column to flow
through the orifice located below the adjacent column.  The particle
fraction values in the region above second of fourth orifice then
oscillate depending on the opened or closed state. This affects the
values of the cumulative particle fraction and eventually the value of
$M$. The period of oscillations in the profile of $M$ is approximately
equal to $T$ and the amplitude is governed by the number of particles
cross flowing within time $T$.  Figure~\ref{fig2}(b) shows a snapshot
at one particular instant for this flow scheme. However, different
cross-flow patterns can be observed during the flow; viz., ``white''
particles (center column) moving through the second and third orifices
and the ``dark red'' (leftmost column) and ``dark blue'' (rightmost
column) moving, respectively, through the second and fourth orifice.
The material collected from each orifice, on an average, now contains
uniformly mixed particles obtained from three different regions.

\begin{figure}
  \centering\includegraphics[width=0.9\linewidth]{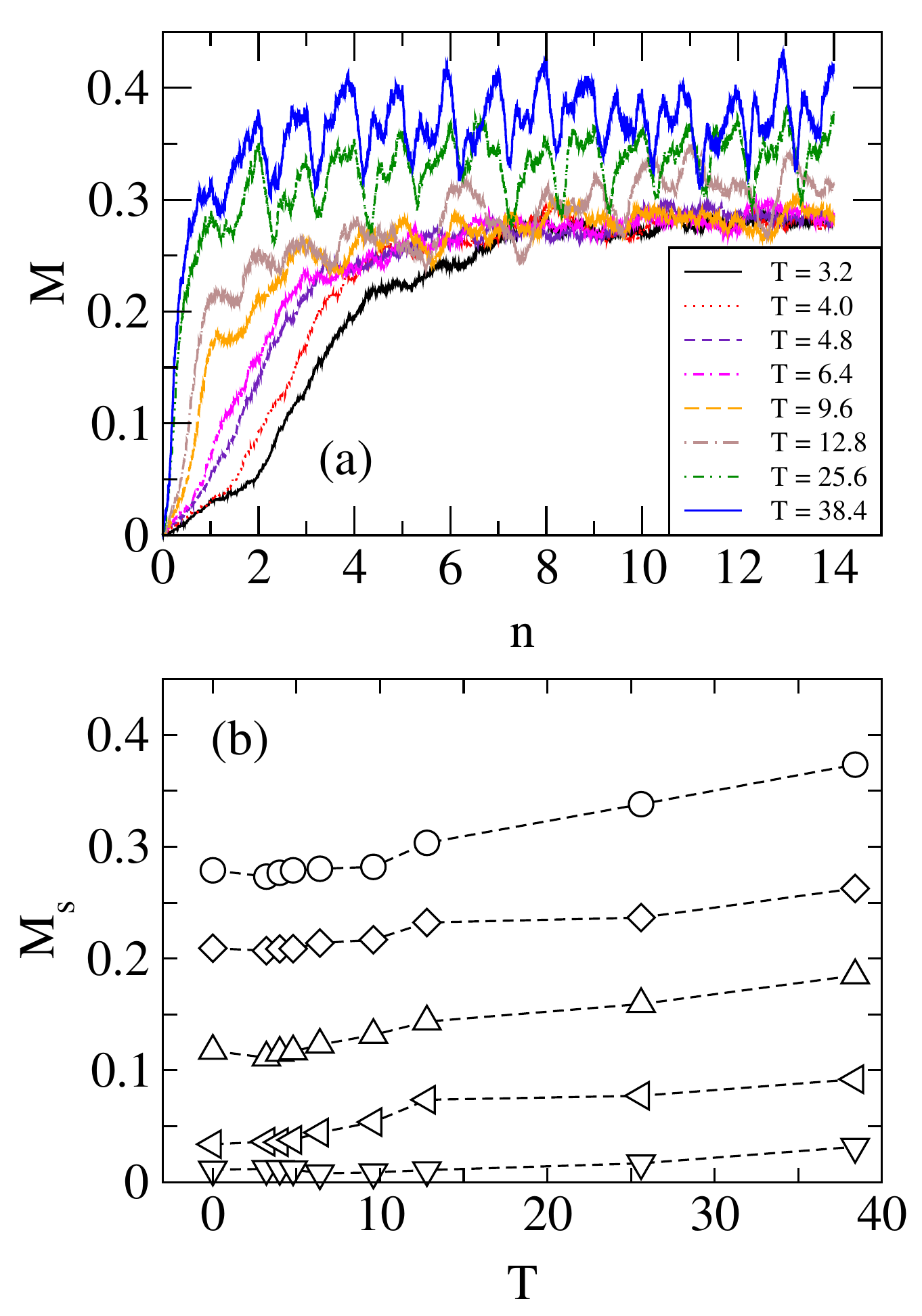}
  \caption{(Color online) (a) Degree of mixing ($M$) plotted against
    $n$. Increased $M$ for increasing values of $T$ at any $n$. (b)
    Final steady state value ($M_{s}$) obtained from the
    values of the polynomial fit (not shown) plotted against
    $T$. Symbols denote values obtained at different heights above the
    orifice: {\large$\circ$} $z = 4d$, {\large$\diamond$} $z = 6d$,
    {\scriptsize$\triangle$} $z = 8d$, {\large$\triangleleft$} $z =
    30d$, {\scriptsize$\nabla$} $z = 110d$.}
  \label{fig3}
\end{figure}

The effect of variation of $T$ on the steady state evolution of mixing
is shown in Fig.~\ref{fig3}(a). For each value of $T$, the number of
instances ($n$) of opening and closing of orifices is kept constant at
fourteen. The total run-time of simulation ($t$), then, increases with
increasing $T$ to achieve the steady state mixing. For higher values
of $T$, fewer instances ($n$) of closing and opening the orifices are
necessary to reach steady state. Furthermore, the values of $M$
increase with increase in $T$ for any value of $n$. For larger values
of $T$, the time available for the cross flow across the regions above
different orifices is greater, leading to higher mixing, hence higher
$M$.  The amplitude of the oscillations of the steady-state continue
to decrease with decreasing $T$. For smaller $T$, the duration of
cross flow is quite small (hence smaller period) and consequently
smaller amplitude (i.e., lesser particles cross flowing during a
particular sequence). For very small values of $T$ ($\approx 0.0$) ,
the system asymptotically approaches the flow and mixing behavior
observed in Fig.~\ref{fig2}(a).

The final steady state value of $M$ obtained from the profiles in
Fig.~\ref{fig3}(a) are shown in Fig.~\ref{fig3}(b) (symbols shown as
open circles). The steady state value for all cases is obtained by
using a tenth-degree-polynomial fit, as described previously.  The
value of $M_{s}$ shows maximum change (by about $15$\%) for $T >
10$. For lower values of $T$, the degree of mixing is nearly the same
as obtained when all orifices are kept open ($T
=0$). Figure~\ref{fig3}(b) also shows the values of $M_{s}$ measured
at different heights above the orifice for varying values of $T$. The
same value of $n$ is needed to achieve the steady state at all
heights, i.e., the steady state is achieved everywhere simultaneously
(profiles not shown). For any value of $T$, the degree of mixing
decreases with increasing height above the orifice.  The dependence on
$T$ is qualitatively the same for all heights up to $30d$, above which
the material moves like a plug and seems unaffected by the
controlled-mechanism protocol employed. The value of $M_{s}$ remains
close to $0.0$, irrespective of the value of $T$ employed.

\begin{figure}
  \centering\includegraphics[width=0.9\linewidth]{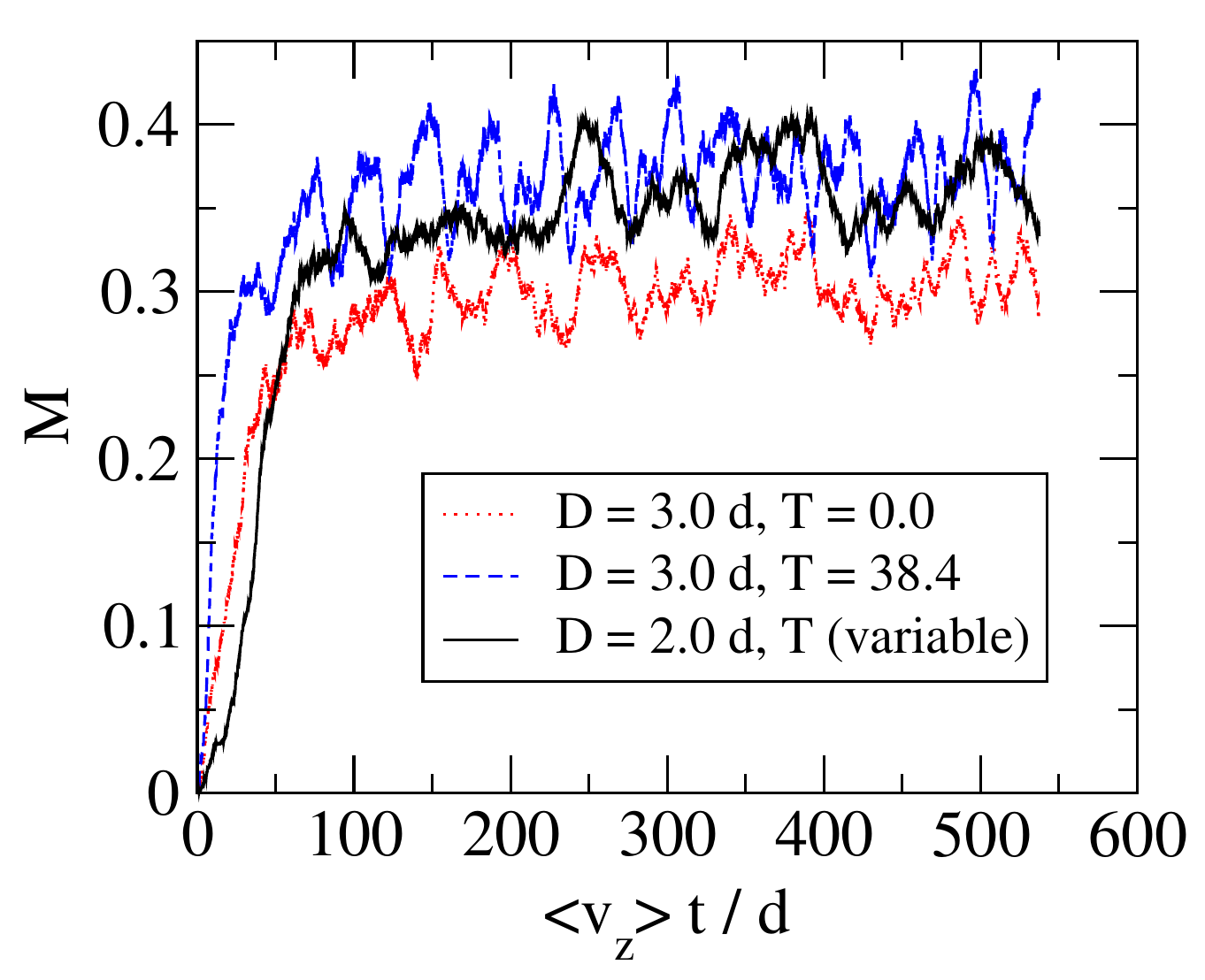}
  \caption{(Color online) Evolution of mixing with time for different
    orifice diameter and $T$. Results for $D = 2.0 d$ are obtained by
    allowing system to choose the $T$ which adjusts to the spontaneous
    jamming and unjamming occurring at different orifices.}
  \label{fig4}
\end{figure}

Next, we compare the mixing behavior for different flowing protocols,
viz., controlled and random mechanisms.  In the former, only the
second and fourth orifices can close and open, and that too with a
definite period, $T$. In the latter case, one or more orifices can jam
or unjam any number of times, leading to variable $T$. The evolution
of mixing for this latter case along with that for the controlled
mechanism with $T = 38.4$ and the base case of $T = 0$ is shown in
Fig.~\ref{fig4}. The total run-time $t$ in simulations is adjusted so
that the mean distance traveled by particles remains the same in all
three cases. The value of $M_{s}$ for the random mechanism is quite
close to that obtained for $T = 38.4$, although it is achieved a bit
slowly. In the initial period, the degree of mixing is even lower than
that for $T = 0.0$. The periodicity of the oscillations are not
defined clearly due to the highly variable $T$ in this protocol. This
protocol, as mentioned earlier, has limited range within which the
interorifice distance and the orifice width can be varied. Increasing
the orifice width ($D$) to $2.2d$ creates flow similar to that for $T
= 0.0$, i.e., there are hardly any instances of flow jamming and the
flow never stops from any of the orifices. While for widths ($D$)
below $2d$ all five orifices tend to jam together, thus stopping the
flow completely, which can be restarted only by external intervention
(remove a particle from the arch or vibrate the system). Over the
simulation run-time ($t$), the total number of particle efflux is the
least for the random mechanism while it is the most for $T = 0.0$.

\subsection{\label{mpass}{Multipass System}}

Here, we present results for a multipass system, i.e., the situation
wherein every particle, initially filled in the silo, is made to
traverse down the entire silo height multiple times. The results are
presented only for the case of random mechanism using $D = 2d$, $w =
15d$, and $H = 80d$. The results for the controlled mechanism are
qualitatively similar, as evident from Fig.~\ref{fig4}, and hence are
not presented. Certain quantitative differences, though, are
discussed at appropriate places.

\begin{figure}
  \centering\includegraphics[width=0.9\linewidth]{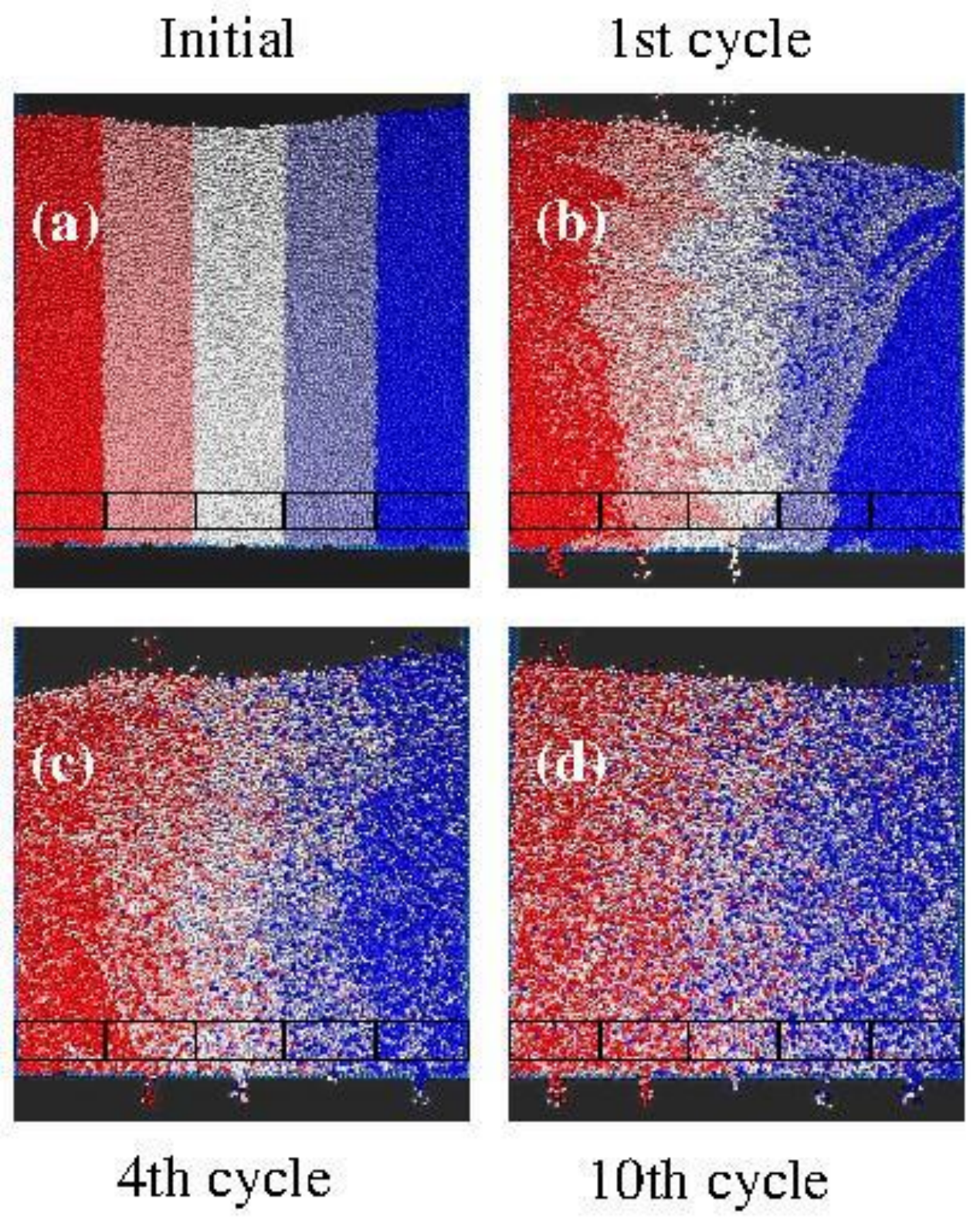}
  \caption{(Color online) Mixing patterns in a silo ($D = 2d$, $w =
    15d$) operated as a multipass system. Opening and closing of any
    orifice is chosen based on random jamming and unjamming
    events. Black boxes at the bottom represent the region in which
    the particle fraction is calculated.}
  \label{fig5}
\end{figure}

\begin{figure}
  \centering\includegraphics[width=0.85\linewidth]{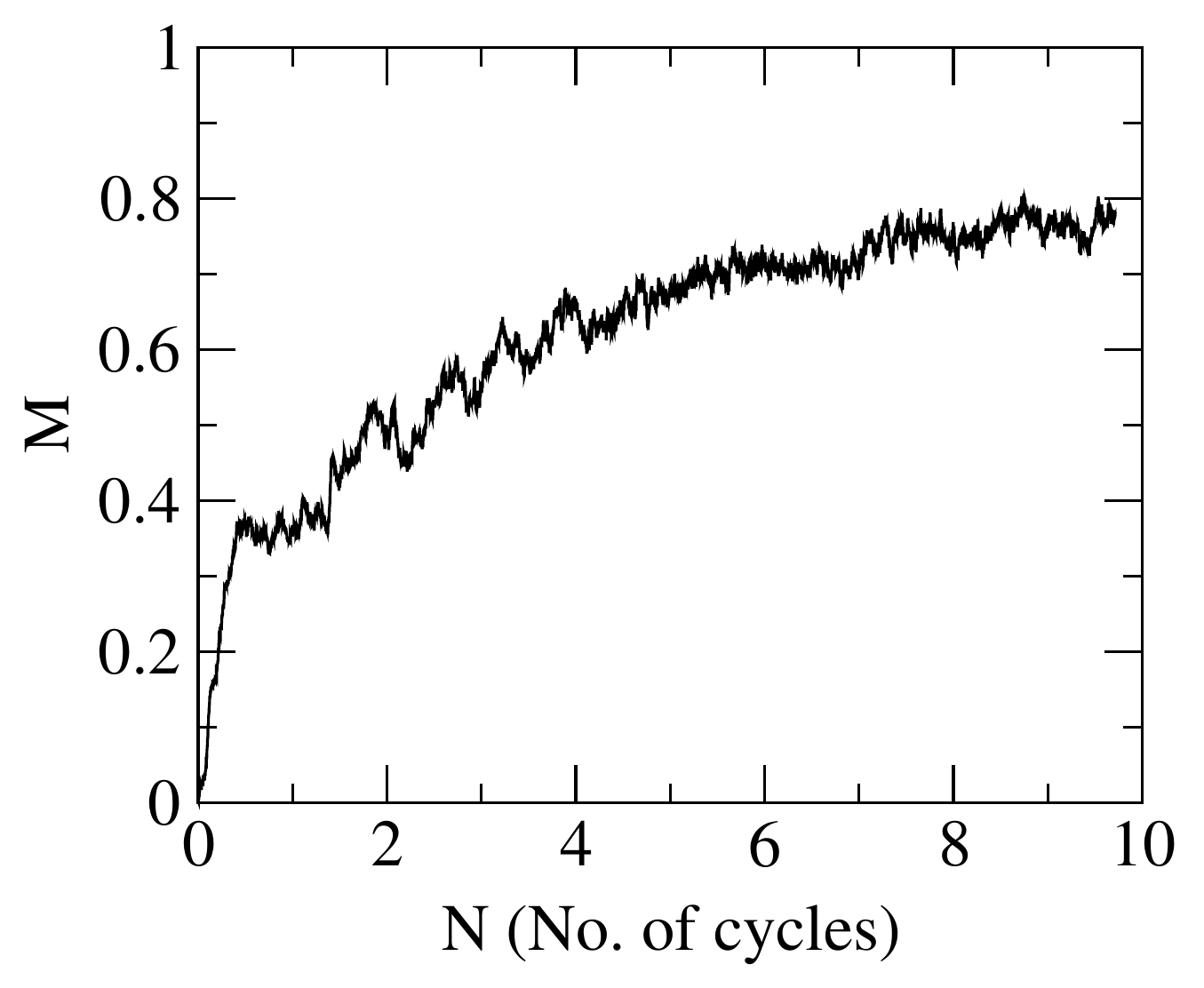}
  \caption{Change in the degree of mixing with flow cycles. The
    maximum value of mixing achievable
    when all colored particles are distributed uniformly across the
    horizontal distance is 1.}
  \label{fig6}
\end{figure}

The mixing patterns are shown in the snapshots in
Fig.~\ref{fig5}(a)-5(d) for the initial state and after different flow
cycles ($N = 1, 4, 10$). One flow cycle represents every particle in
the silo traversing the entire height ($H$) once and would
approximately correspond to $v_{z} \Delta t/d = 80d$.  The movie
showing the entire simulation run is provided as supplementary
material.  From the images of the initial state and that after the
$10$th cycle, it seems as if significant mixing has taken place across
the width of the silo. This is confirmed by the value of the degree of
mixing $M$ calculated in the region covered by the black boxes of same
size and located at same height as used in a single-pass system.  The
final steady state value of $M$ is quite high, around $0.75$ (see
Fig.~\ref{fig6}), but still lower than $1.0$ which represents complete
mixing. It seems that complete mixing will be achievable after an
infinite number of cycles, as evident from the very slow rate of
change in the value of $M$ after eight cycles.

The overall mixing behavior in the multi-pass system can arise out of
three different mechanisms. The first can be due to the impact of the
particles hitting the free surface and then bouncing horizontally in
the nearby regions. This does not seem to be the case, as observed
from the particle trajectories (not shown over here). The particles
simply fall onto the free surface and start moving downward showing
very little lateral movement due to impact. The second mechanism is
due to slope formation at the free surface [see Fig.~\ref{fig5}(b)]
leading to an avalanche of particles, thereby transferring the
particles to nearby regions, thus causing horizontal spreading. This
is evident from the horn of ``light-blue'' particles (second column
from right wall) spreading up to the silo wall, as seen clearly in
Fig.~\ref{fig5}(b). The creation of local slope is due to the certain
columns remaining stationary due to jammed orifice below them and all
the particles exiting from other orifices and piling up at the
corresponding free surface above. The third mechanism is due to the
particles in a region above a certain orifice flowing out from the
nearby orifice due to jamming of the orifice below it and then
re-entering the silo in the zone from where it exited instead of where
it started in the previous cycle. This is the same mechanism
responsible for the profiles of $M$ shown in Fig.~\ref{fig4} but is
applied every cycle. The relative dominance of these two mechanisms is
difficult to ascertain quantitatively. The second mechanism is
expected to be dominant when orifices are jammed for an extended
period of time causing the material to pile up for avalanching at the
free surface. In the scenario where there is rapid jamming and
unjamming, the third mechanism is expected to dominate.

The repeated application of the second and third mechanisms per cycle
eventually causes a particle to drift significantly from its original
horizontal position at the start of the simulation run. Even a few of
the particles from the regions in the vicinity of the wall manage to
reach the opposite side wall of the silo. However, this number is very
restricted as the bulk of the particles in the extreme columns do not
show significant horizontal drift even after ten cycles, thus
preventing complete mixing. To some extent, this is also due to the
limitation enforced on the particles in the first and fifth columns to
move only in one direction, the other being blocked by the
sidewall. The final mixed state of the silo shown in
Fig.~\ref{fig5}(d) is observed to be independent of the height above
the orifice, i.e., the same horizontal variation is observed at all
heights.

If the controlled mechanism (with highest $T$) is employed instead of
the random mechanism in a multipass mode, the degree of mixing
achieved at all times (cycles) is about $10$\% lower. In the latter
case, every orifice closes at least once with few others closing
multiple times and for any duration and, furthermore there are certain
instances of two orifices being closed simultaneously. The overall
result is to enable particles from all regions of the silo to undergo
the mixing process, which results in a more homogeneous mixture. While
for the controlled mechanism, only one (second or fourth) orifice is
closed at any point of time and that, too, for a fixed time
period. Thus, only certain specific regions (mostly the three central
columns) of the silo get involved in mixing, resulting in less
homogeneous mixing. It is obvious that the controlled mechanism, if
modified to close and open all orifices for a varied time, would
result in improved mixing of the same order, if not better, as for the
case of the random mechanism. However, the scheme employing random
mechanism seems more easily realizable in practice.

We would like to note that the phenomena of horizontal shifting of the
particle positions at every flow cycle has striking similarities with
those encountered in a quasi-two-dimensional horizontal rotating
cylinder~\cite{hill99a,hill99b,kha99}, which show chaos due to changing
streamline positions. In a rotating cylinder, this streamline shifting
is achieved due to flowing layer thickness fluctuations induced by
varying rotation rate or using cylinders of noncircular
crosssection. Here, this shifting is achieved by the inherent
randomness in the opening and closing of the orifices. It is not clear
at the moment if the eventual mixing behavior is a result of possible
chaos achieved in the system. This would require a more extensive study
using a model which can exhibit flow discontinuities in a jamming and
unjamming scenario. Nevertheless, it is quite interesting to know
about the existence of possible chaos in a silo system which arises
out of system randomness.
 
\section{\label{concl}Summary}

To summarize, our study shows that the mixing behavior of particles
draining down a silo can be enhanced by using multiple orifices and
clever manipulation of the choice of orifices which can close and open
frequently. The closing of an orifice leads to the crossflow of
particles between different regions, thus improving mixing.  More
importantly, the overall scheme employed eliminates the stagnation
zones in the system while ensuring a nearly uniform particle fraction
across the silo width, a feature missing in the currently used
single-orifice silos. The desired level of mixing can either be
obtained by controlled opening and closing of specific orifices or
allowing the system to choose on its own using the inherent random
jamming and unjamming behavior. Different schemes employed provide a
realistically usable system in practice: operating the silo with a
smaller height in a single or multipass mode by repeated recirculation
of the particles emanating from the exit orifices. Operating over
longer times would ensure a highly uniform mixture of particles across
the entire silo, a feature which is not that easy to achieve even in a
more familiar rotating cylinder system. The results obtained by
subjecting particles to repeated horizontal crossflow show promise for
achieving a chaos in the system arising out of the inherent system
randomness.  While the overall results show promise for dry
particulate systems, their applicability to a more complicated and
practically relevant cohesive granular system remains to be explored
further. In that case, the orifice size and interorifice distance may
have to be varied, perhaps, to account for a large effective particle
size due to formation of clusters of individual cohesive particles.
More interesting would be to study the mixing behavior of mixtures of
particles of varying size and/or density which would need varying
orifice sizes and interorifice distances in the same system.

\begin{acknowledgements} 
  We thank Mayuresh Kulkarni for carrying out preliminary
  experiments and the funding from Department of Science and
  Technology, India, Grants No. $SR/S3/CE/037/2009$ and
  $SR/S3/CE/0044/2010$.
\end{acknowledgements} 

\bibliography{mix}

\end{document}